\documentclass[graybox,envcountchap,sectrefs]{svmono}

\usepackage{mathptmx}
\usepackage{helvet}
\usepackage{courier}
\usepackage{type1cm}         

\usepackage{makeidx}         
\usepackage{graphicx}        
\usepackage{multicol}        
\usepackage[bottom]{footmisc}
\usepackage{amsmath}

\DeclareMathOperator*{\JLUminimize}{\text{minimize}}

\makeindex             

\begin{document}

\author{Author name(s)}
\title{Book title}
\subtitle{-- Monograph --}



\tableofcontents

\mainmatter
\newcommand{\BE}{\begin{equation}}
\newcommand{\EE}{\end{equation}}
\newcommand{\BA}{\begin{eqnarray}}
\newcommand{\EA}{\end{eqnarray}}
\newcommand{\curl}{\nabla\times}
\newcommand{\minimize}[1]{\JLUminimize_{#1}\;&}
\newcommand{\subto}{\text{subject to}\;&}

\newcommand{\myfig}[2]{\begin{figure}[!h]\includegraphics[width=0.95\textwidth]{fig/#1}\caption{#2}\label{fig:#1}\end{figure}}

\newcommand{\BI}{\begin{itemize}\item}
\renewcommand{\I}{\item}
\newcommand{\EI}{\end{itemize}}

\newcommand{\ER}[1]{\eqref{eq:#1}}
\newcommand{\SR}[1]{Section~\ref{sec:#1}}
\newcommand{\sR}[1]{section~\ref{sec:#1}}
\newcommand{\FR}[1]{Figure~\ref{fig:#1}}
\newcommand{\fR}[1]{figure~\ref{fig:#1}}
\chapter{Objective-First Nanophotonic Design}
\label{intro}

\abstract{We introduce an ``objective-first'' strategy 
    for designing nanophotonic devices,
    and we demonstrate the design of nanophotonic
    coupler, cloak, and mimic devices.}

Our initial foray into design methods for nanophotonic devices
    began with a very simple and naive question: 
    could we make an inverse solver which,
    when given the electromagnetic fields we desire,
    returns the nanophotonic structure that will produce them \cite{opex3}?
In other words, since we already know how to solve for $E$ and $H$ 
    in Maxwell's equations, why can't we solve for $\epsilon$
    or even $\mu$ instead?

Not surprisingly, it did not take us long to find that such
    a simple strategy would inevitably run into many problems.

Over the subsequent years, we were able to come up with a better solution,
    which we call an ``objective-first'' strategy for nanophotonic design,
    and which we present in this chapter.
Although it is much more advanced than our original idea,
    objective-first design still carries the same fundamental concept,
    which is to specify the electromagnetic fields, and then
    to solve for a structure to produce them.
    
In this chapter, we present the simple theoretical underpinnings of 
    objective-first design in the first two sections,
    and then show examples of the method in action in the rest of the chapter.
We also include the source code that was used to generate
    all results presented herein \cite{code}.

\section{The electromagnetic wave equation}
In this section, we outline the wave equation
    that is central to the application of our method,
    with the end-result being to show that it is 
    separably linear (bi-linear) in the field and structure variables.
We do this by first formulating this wave equation in the language of physics,
    and then discretizing it in order to achieve numerical solutions.
We then show how one can not only obtain the solution for the field,
    but also obtain the solution for the structure using
    simple, standard numerical tools.

\subsection{Physics formulation}
First, let's derive our wave equation,
    starting with the differential form of Maxwell's equations, 
\BA \curl E = - \mu_0 \frac{\partial H}{\partial t} \\
    \curl H = J + \epsilon \frac{\partial E}{\partial t}, \EA
    where $E$, $H$, and $J$ are 
    the electric, magnetic and electric current
    vector fields, respectively,
    $\epsilon$ is the permittivity
    and $\mu_0$ is the permeability, which we assume to be 
    that of vacuum everywhere.

Assuming the time dependence $\exp(-i \omega t)$, 
    where $\omega$ is the angular frequency,
    these become
\BA \curl E = - i \mu_0 \omega H \\
    \curl H = J + i \epsilon \omega E, \label{eq:H2E} \EA
    which we can combine to form our (time-harmonic) wave equation,
\BE \curl \epsilon^{-1} \curl H - \mu_0 \omega^2 H = \curl \epsilon^{-1} J. 
    \label{eq:wave} \EE

In this chapter, we are going to only consider the two-dimensional form
    of this equation, and specifically 
    the two-dimensional transverse electric (TE) mode\cite{taflove}.
In this case \ER{wave} is simplified 
    because only the $z$-component of $H$ is non-zero.

Nevertheless,  a single equation \ER{wave},
    represents all the physics which we take into account in this chapter.

\subsection{Numerical formulation}
On top of the analytical formulation of the wave equation \ER{wave}
    we will now add a numerical, or discretized, formulation.
This will be needed in order to solve for arbitrary structures
    for which there are not analytical solutions.

The salient step in order to do so
    is to the use of the Yee grid\cite{yee}, 
    which allows us to easily define the curl $(\curl)$
    operators in \ER{wave}. 
Since both the individual curl operators and the equation as a whole
    is linear in $H$, it naturally follows to formulate \ER{wave},
    with a change of variables, as
\BE A(p)x = b(p), \label{eq:Ab} \EE 
    where $H \to x$, $\epsilon^{-1} \to p$; 
    and where
\BE A(p) = \curl \epsilon^{-1} \curl - \mu_0 \omega^2 \EE 
    and
\BE b(p) = \curl \epsilon^{-1} J. \EE
Note that our use of $A(p)$ and $b(p)$ instead of $A$ and $b$
    simply serves to clarify the dependence
    of both $A$ and $b$ to $p$.

Apart from using the Yee grid, the only other salient implementation detail
    is the use of
    stretched-coordinate perfectly matched layers \cite{pml}
    where necessary, in order to prevent unwanted reflections
    at the boundaries of the simulation domain.
The effect of such layers is to modify the curl operators,
    although their linear property is still maintained.

\subsection{Solving for $H$}
With our numerical formulation, we can now solve for the $H$-field
    (the $E$-field can be computed from the $H$-field using \ER{H2E})
    by applying general linear algebra solvers to \ER{Ab}.
Recall that since we have chosen a time-harmonic formulation,
    solving for $x$ in \ER{Ab} is actually performing what is simply known as
    a time-harmonic or a finite-difference frequency-domain (FDFD) simulation\cite{shin}. 
Furthermore, since we have limited ourselves to the two-dimensional case,
    \ER{Ab} is easily solved using the standard sparse solver
    included in Matlab on a single desktop computer.

We call the routine that solves for $x$ in \ER{Ab} given $p$ a field-solver,
    or a simulator.

\subsection{Solving for $\epsilon^{-1}$}
    After having built a field-solver or simulator
    (which finds $x$ given $p$) for our wave equation,
    the next step is to build a structure-solver for it.
In other words, we need to be able to solve for $p$ given $x$.

To do so, we return to \ER{wave}
    and remark that 
    $\epsilon^{-1} (\curl H) = (\curl H) \epsilon^{-1}$ and
    $\epsilon^{-1} J = J \epsilon^{-1}$ 
    since scalar multiplication is commutative.
This allows us to rearrange \ER{wave} as
\BE \curl (\curl H) \epsilon^{-1} - \curl J \epsilon^{-1}  = \mu_0 \omega^2 H  \EE
which we now write as 
\BE B(x)p = d(x), \label{eq:Bd} \EE 
    where
\BE B(x) = \curl (\curl H) - \curl J\EE
    and 
\BE d(x)  = \mu_0 \omega^2 H.  \EE

With this extremely simple trick,
    we have shown that we can seemingly solve for $p$ given $x$
    with approximately the same ease as solving for $x$ given $p$!
We see this because the dimensions and complexity of $B(x)$ are basically
    equivalent to that of $A(p)$,
    and this implies that the same simple tools used in our field-solver
    should be applicable to solving \ER{Bd}.
This is indeed what we find, although the later addition of constraints on $p$
    will require the use of more powerful (but just as dependable) numerical tools.

\subsection{Bi-linearity of the wave equation}
Although additional mathematical machinery must still be added
    in order to get a useful design tool,
    we have shown so far that the wave equation is 
    separately linear or in $x$ and $p$ (i.e. bilinear).
Namely,
\BE A(p)x-b(p) = B(x)p - d(x). \label{eq:bilinear} \EE
In other words, fixing $p$ makes solving the wave equation for $x$
    a linear problem, and vice versa.
Note that the joint problem,
    where both $x$ and $p$ are allowed to vary,
    is not linear.

The bi-linearity of the wave equation
    is \emph{absolutely fundamental} in our objective-first strategy
    because it relies on the fact
    that, although simultaneously solving for $x$ and $p$ is very difficult,
    we already know how to solve linear systems ($x$ and $p$ separately) well.
In fact, it is this very property which forms the natural division of labor
    which our objective-first method exploits.

\section{The objective-first design problem}
We now describe the remaining machinery used in the objective-first method,
    in addition to the field-solver and the structure-solver,
    as previously outlined.
Specifically, we introduce the idea of a design objective and a physics residual,
    and we reference the mathematical notion of convexity
    in order to motivate the need to divide the 
    objective-first problem into two separately convex sub-problems.

\subsection{Design objectives} \label{sec:desobj}
A design objective, $f(x)$, is simply defined as
    a function we wish to be minimal 
    for the design to be produced.

For instance, in the design of a device
    which must transmit efficiently into a particular mode,
    we could choose $f(x)$ to be the negative power flow into that mode.
Or, if the device was to be a low-loss resonator,
    we could choose $f(x)$ to be the amount of power leaking 
    out of the device.

In general, there are multiple choices of $f(x)$
    which can be used to describe the same objective.
For example, $f(x)$ for a transmissive device 
    may not only be the negative power transmitted into the desired output mode,
    but it could also be the amount of power lost to other modes,
    or even the error in the field values at the output port
    relative to the field values needed for perfect transmission.
These design objectives are equivalent in the sense that, if minimized, 
    all would produce structures with good performance.
At the same time, we must consider that the computational cost and complexity
    of using one $f(x)$ over another may indeed vary greatly.

\subsection{Convexity}
Before formulating the design problem,
    we would like to add a note regarding the complexity of various 
    optimization problems.

Specifically, we want to introduce the notion of \emph{convexity}\cite{boyd}
    and to note the difference between problems that
    are convex and those which are not.
The difference is simply this:
    convex problems have a single optimum point
    (only one local optimum, which is therefore the global optimum)
    which we can reliably find using existing numerical software,
    whereas non-convex problems typically have multiple optima 
    and are thus much more difficult to reliably solve.

That a convex problem can be reliably solved, in this case, 
    means that regardless of the starting guess,
    convex optimization software will 
    always arrive at the globally optimal solution
    and will be able to numerically prove global optimality as well.
Thus, the advantage in formulating a design problem 
    in terms of convex optimization problems
    is to eliminate both the need to circumvent local optima and 
    any notion of randomness.

On a practical level, there exist mature convex optimization software packages
    among which is CVX, a convex optimization package written for Matlab\cite{cvx}, 
    which we use for the examples in this chapter.

\subsection{Typical design formulation}
We now examine the typical, and most straightforward formulation
    of the design problem,
    in order to relate and contrast it to the objective-first formulation.
The design problem for a physical structure is typically formulated as
\BA \minimize{x,p} f(x) \label{eq:typform} \\
    \subto A(p)x - b(p) = 0, \notag \EA
    which states that we would like to vary $x$ and $p$ simultaneously
    in order to decrease $f(x)$
    while always satisfying physics (e.g. the electromagnetic wave equation).

Since solving \ER{typform} is quite difficult in the general sense
    (simultaneously varying $x$ and $p$ makes the problem non-convex),
    traditional approaches have relied on either brute-force parameter search,
    or a gradient-descent method utilizing first-order derivatives.
In the gradient-descent case, solving \ER{typform} results in the
    well-known adjoint optimization method\cite{miller}. 

\subsection{Objective-first design formulation}
In contrast with the typical formulation, the objective-first formulation 
    simply switches the roles of 
    the wave equation and the design objective with one another,
\BA \minimize{x,p} \| A(p) x - b(p) \|^2 \label{eq:ob1:1} \\
    \subto f(x) = f_\text{ideal}. \label{eq:ob1:2} \EA
Although such a switch may seem trivial,
    and even silly at first,
    we show that it fundamentally changes the nature of the design problem
    and actually gains us advantages in our efforts at finding a solution.

This first fundamental change, as seen from \ER{ob1:1},
    is that we allow for non-zero residual in the electromagnetic wave equation.
This literally means that we allow for \emph{non-physical} $x$ and $p$,
    since $A(p) x - b(p) \ne 0$ is permissible.
And since $A(p) x - b(p)$ can now be a non-zero entity, 
    we choose to call it the \emph{physics residual}.  
The second fundamental change
    is that we always force the device to exhibit ideal performance,
    as seen from \ER{ob1:2}.
This, of course, ties in very closely with \ER{ob1:1} since
    ideal performance is usually not obtainable unless one
    allows for some measure of error in the underlying physics
    (non-zero physics residual).
As such, our strategy will be to vary $x$ and $p$
    in order to decrease the physics residual \ER{ob1:1} to zero,
    while always maintaining ideal performance.

The primary advantage in the objective-first formulation is that,
    although the full problem is still non-convex,
    it allows us to form two convex sub-problems, as we outline below.
In contrast to an adjoint method approach,
    in doing we can still access information regarding
    second-order derivatives, which greatly speeds up finding a solution.
An additional advantage is that our insistence
    that ideal performance be always attained
    provides a mechanism
    which can potentially ``override'' local optima 
    in the optimization process.

To this end we have found that such a strategy
    actually allows us to design very unintuitive devices
    which exhibit very good performance,
    even when starting from completely non-functional initial guesses.
Furthermore, we have found this to be true
    even true when the physics residual is never brought to exactly zero.


In practice, we add an additional constraint to the original formulation, \cite{opex1}
    which is to set hard-limits on the allowable values of $p$,
    namely $p_0 \le p \le p_1$.
This is actually a relaxation of the ideal constraint,
    which would be to allow $p$ to only have discrete values,
    $p \in p_0, p_1$,
    but such a constraint would be essentially force us to only
    be able to perform brute force trial-and-error.

Our objective-first formulation is thus,
\BA \minimize{x,p} \| A(p) x - b(p) \|^2 \notag \\
    \subto f(x) = f_\text{ideal} \label{eq:ob1} \\
        & p_0 \le p \le p_1, \notag \EA
    which is still non-convex, but can be broken down into 
    two convex sub-problems, 
    the motivation being that each of these will be 
    able to be easily and reliably solved.

\subsection{Field sub-problem}
The first of these is the field sub-problem, 
    which simply involves fixing $p$ and independently optimizing $x$,
\BA \minimize{x} \| A(p) x - b(p) \|^2 \label{eq:Fsub} \\
    \subto f(x) = f_\text{ideal}. \notag \EA
This problem is convex, and actually quadratic,
    which means that it can even be solved 
    using standard numerical tools, in the same way 
    as a simple least-squares problem.

The field sub-problem can be thought of as an update to $x$ (field)
    where we try to ``fit'' the electromagnetic fields to the structure ($p$).
Of course, if it were not for the hard-constraint on the design objective,
    the field sub-problem would be able to perfectly fit $x$ to $p$.
This, it turns out, would exactly be a simulation.

\subsection{Structure sub-problem}
The second sub-problem is formulated by fixing $x$ and
    independently optimizing $p$.
At the same time, we use the bi-linearity property
    of the physics residual from \ER{bilinear}
    to rewrite the problem in a way that makes
    its convexity explicit,
\BA \minimize{p} \| B(x) p - d(x) \|^2 \label{eq:Ssub} \\
    \subto p_0 \le p \le p_1. \notag \EA
The structure sub-problem is also convex, but not quadratic because of the 
    inequality constraints on $p$.
However, use of the CVX package still allows us to obtain the result
    quickly and reliably.

Note that in an analogous fashion to the field sub-problem,
    the structure sub-problem attempts to fit $p$ to $x$,
    and is prevented from perfectly doing so by its own constraint.

Because neither sub-problem is capable of completely reducing the physics residual
    to zero, they must be used in an iterative manner in order to
    gradually decrease the physics residual.
To this end, we employ the alternating directions optimization method.

\subsection{Alternating directions}
We use a simple alternating directions scheme 
    to piece together \ER{Fsub} and \ER{Ssub},
    which is to say that we simply
    alternately solve each and continue until we reach some stopping point,
    normally measured by how much the physics residual has decreased.
\BA \text{Loop:} & & \notag \\ 
& \minimize{x} \| A(p) x - b(p) \|^2 \notag \\
 &    \subto f(x) = f_\text{ideal}. \notag \\ 
    \\
& \minimize{p} \| B(x) p - d(x) \|^2 \notag \\
    & \subto p_0 \le p \le p_1. \notag \EA
The alternating directions scheme is extremely simple
    and does not require additional processing
    of $x$ or $p$ outside of the two sub-problems,
    nor does it require the use of auxiliary variables.

The advantage of such the alternating directions method
    is that the physics residual is guaranteed to
    monotonically decrease with every iteration,
    which is useful in that no safeguards
    are needed to guard against ``rogue'' steps
    in the optimization procedure.
Note that this robustness stems from the fact that,
    among other things,
    each sub-problem does not rely on previous values of 
    the variable which is being optimized,
    but only on the variable which is held constant.

The disadvantage of such a simple scheme is that 
    the convergence is quite slow,
    although we have found it to be sufficient in our cases.
Related methods, such as the Alternating Directions Method of Multipliers\cite{admm}, 
    exhibit far better convergence.

\section{Waveguide coupler design}\label{sec:wg}
We first apply the objective-first formulation
    with the alternating directions algorithm
    to the design of nanophotonic waveguide couplers
    in two dimensions,
    where our goal is to couple light from
    a single input waveguide mode
    to a single output waveguide mode
    with as close to unity efficiency as possible.
We would also like to allow the user to choose arbitrary
    input and output waveguides,
    as well as to select 
    arbitrary modes within those waveguides
    (as opposed to allowing only the fundamental mode, for example).

This problem is very general and, in essence,
    encompasses the design of all linear nanophotonic components,
    because the function or performance of all such components
    is simply to convert a defined set of input modes
    into a defined set of output modes.
Such a broad, general problem is ideally suited for 
    an objective-first strategy,
    since no approximations or simplifications
    of the electromagnetic fields are required;
    we only make the simplification of working in two dimensions
    (transverse magnetic mode)
    and dealing only with a single input and output mode.

\subsection{Choice of design objective}
As mentioned in \sR{desobj} multiple equivalent choices
    of design objective exist which should allow one
    to achieve the same device performance;
    however, we will choose, for generality, the following design objective,
\BE f(x) = \begin{cases}
        x - x_\text{perfect} & \text{at boundary}, \\
        0 & \text{elsewhere},
        \end{cases} \EE
That is, $f(x)$ simply selects the outermost values of the field
    in the design space
    and compares them to values of a perfect device.

Furthermore, we choose $f_\text{ideal} = 0$ so that
    when placed into the objective-first problem \ER{ob1},
    this will result in fixing the boundary values of the field
    at the edge of the design space
    to those of an ideal device,
    as shown in \fR{wg/intro}.
In this case, we choose such an ideal device
    to have perfect (unity) coupling efficiency,
and these ideal fields are simply obtained by using
    the input and output mode profiles at the corresponding ports
    and using values of zero at the remaining ports.

\myfig{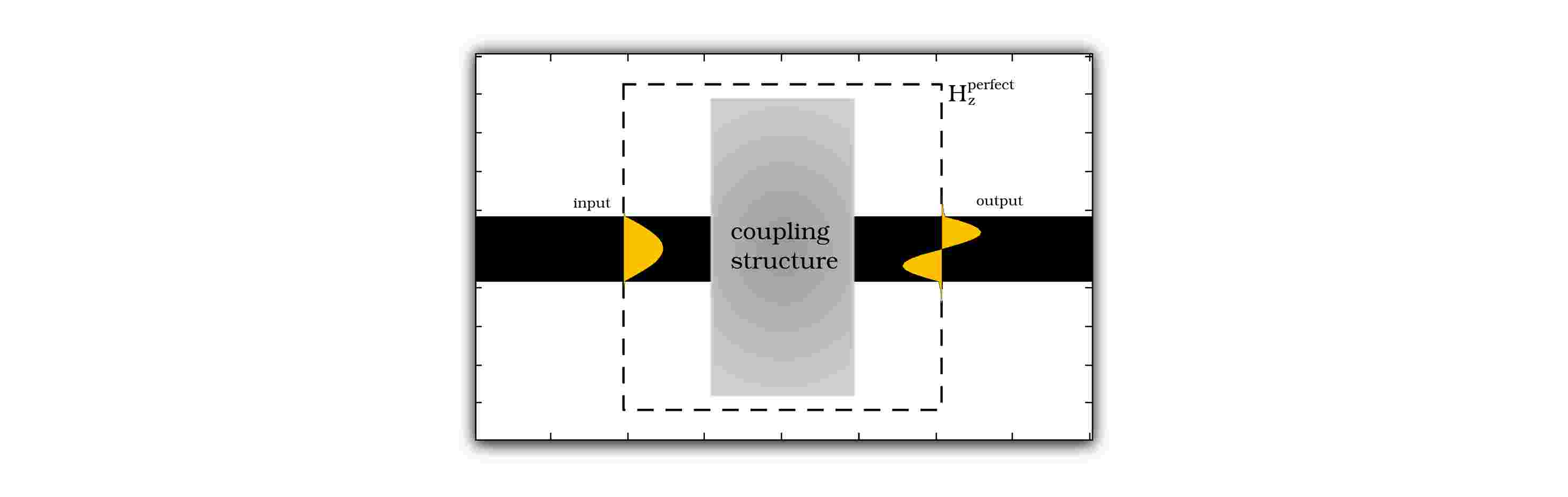}{Formulation of the design objective.}

Such a design objective is general
    in the sense that the boundary values of the device 
    contain all the information necessary to determine
    how the device will interact with its environment,
    when excited with the input mode in question.
In other words,
    we only need to know the boundary field values,
    and not the interior field values to determine 
    the performance of the device;
    and thus, it would be conceivable that such a scheme
    might be generally applied to linear nanophotonic devices beyond 
    just waveguide mode couplers.

In our case,
    we only need to know the value of $H_z$ and 
    its derivative along the normal direction, $\partial H_z / \partial n$,
    along the design boundary
    in order to completely characterize its performance.
Alternatively,
    one can, of course, use the outermost two layers of the $H_z$
    instead of calculating a spatial derivative.

\subsection{Application of the objective-first strategy}
Having chosen our design objective we apply
    alternating directions to \ER{ob1} which 
    results in solving the following two sub-problems iteratively:
\BA \minimize{x} \| A(p) x - b(p) \|^2 \\
    \subto x = x_\text{perfect} \text{, at boundary} \notag \EA
\BA \minimize{p} \| B(x) p - d(x) \|^2 \\
    \subto p_0 \le p \le p_1 \notag \EA

For the results throughout this chapter, 
    we uniformily choose $p_0 = 1/12.25$ and $p_0 = 1$,
    corresponding to $\epsilon^{-1}$ of silicon and air respectively.
Additionally, since a starting value for $p$ is initially required,
    we always choose to use a uniform value of $p = 1/9$ 
    across the entire design space.
There is nothing really unique about such a choice,
    although we have noticed that initial value of $p$ near 1 
    often result in poor designs.
Note, that, unlike $p$, we do not require an initial guess for $x$.

The only other significant value that needs to be set initially
    is the frequency, or wavelength of light.
We use free space wavelengths in the range of 25 to 63 grid points for
    the results in this chapter.

Lastly, for all the examples presented in the chapter,
    we run the alternating directions algorithm for 400 iterations.
Although we do not present the convergence results here,
    such information can be obtained by inspecting the source code\cite{code}.

\subsection{Coupling to a wide, low-index waveguide}
As a first example, we design a coupler from 
    the fundamental mode of a narrow, high-index waveguide
    to the fundamental mode of a wide, low-index waveguide.
Such a coupler would be useful for coupling from 
    an on-chip nanophotonic waveguide to
    an off-chip fiber for example.

\myfig{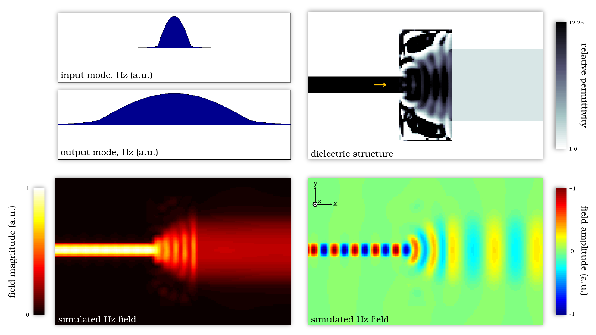}{Coupler to a wide low-index waveguide.
            Efficiency: 99.8\%, 
            device footprint: $36 \times 76$ grid points, 
            wavelength: 42 grid points.}

The input and output mode profiles used as
    the ideal fields are shown in the upper-left corner of \fR{wg/1}.
The final structure is shown in the upper right plot, and
    the simulated $H_z$ fields,
    under excitation of the input mode in this final structure,
    are shown in the bottom plots.

\FR{wg/1} then shows that the design structure has nearly unity efficiency
    and converts between the input and output modes
    within a very small footprint.

\subsection{Mode converter}
In addition to coupling to a low-index waveguide,
    we show that we can successfully apply the objective-first method
    to convert between modes of a waveguide.
We do this by simply selecting the output mode 
    in the design objective to be the second-order waveguide mode,
    as seen in \fR{wg/2}.
\myfig{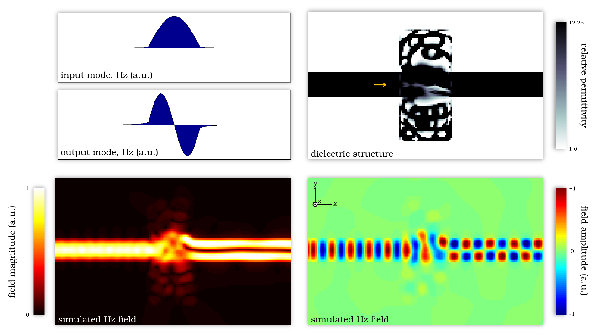}{Mode converter.
            Efficiency: 98.0\%, 
            device footprint: $36 \times 76$ grid points, 
            wavelength: 42 grid points.}

Note that the design of this coupler is made challenging
    because of the opposite symmetries of the input and output modes.
Moreover, because our initial structure is symmetric,
    we initially have exactly 0\% efficiency to begin with.
Fortunately, the objective-first method can still design
    an efficient coupler in this case as well.

\subsection{Coupling to an air-core waveguide mode}
We can then continue to elucidate the generality of our method
    by coupling between waveguides which confine light 
    in completely different ways.
\myfig{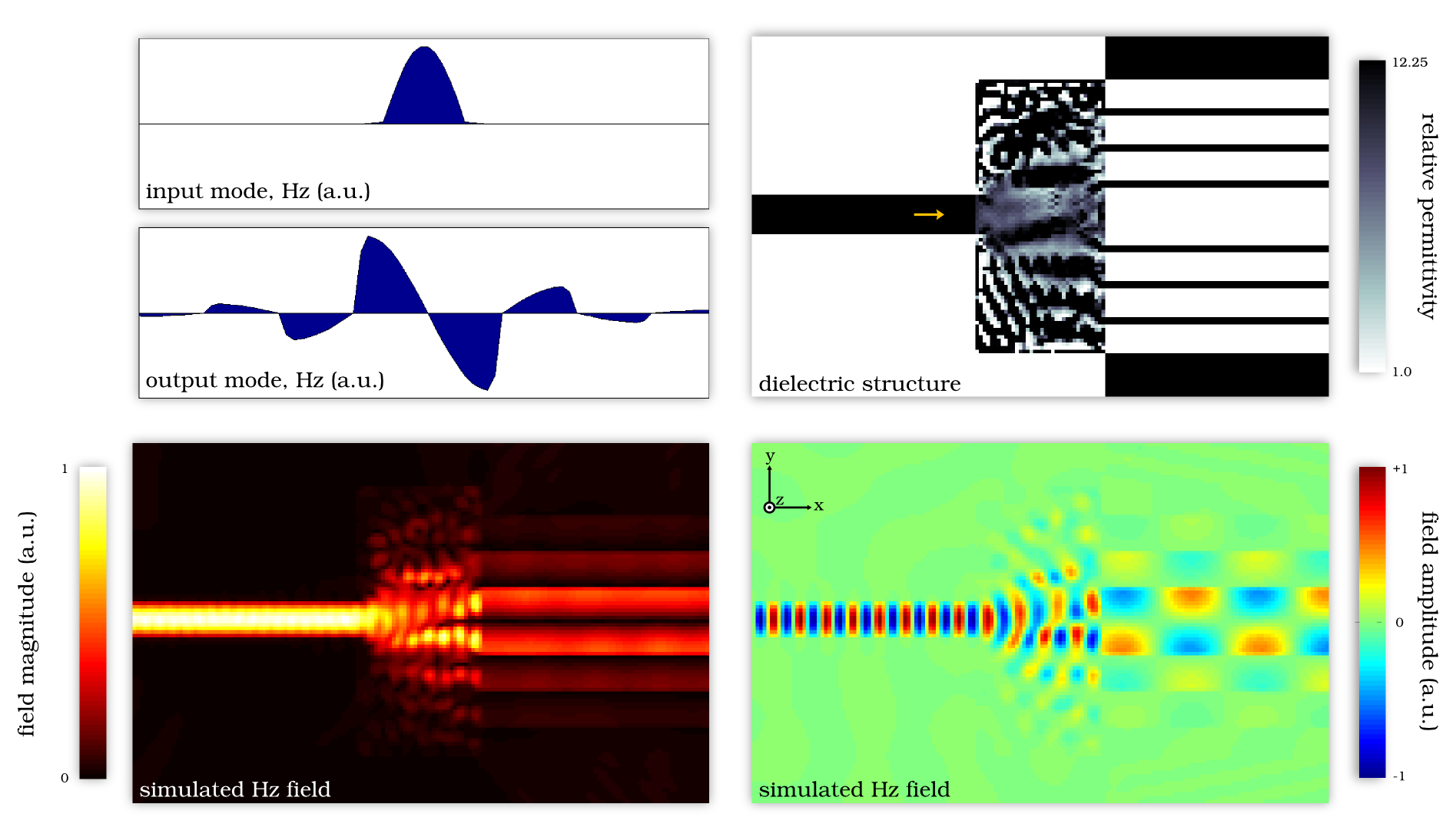}{Coupler to a wide low-index waveguide.
            Efficiency: 98.9\%, 
            device footprint: $36 \times 76$ grid points, 
            wavelength: 25 grid points.}

\FR{wg/3} shows  a high-efficiency coupling device between 
    an index-guided input waveguide and
    a ``air-core'' output waveguide, 
    in which the waveguiding effect is achieved using distributed Bragg reflection
    (instead of total internal reflection as in the input waveguide).

\subsection{Coupling to a metal-insulator-metal waveguide }
Additionally, our design method can also generate couplers
    between different material systems such as 
    between dielectric and metallic (plasmonic) waveguides,
    as shown in \fR{wg/4}.
\myfig{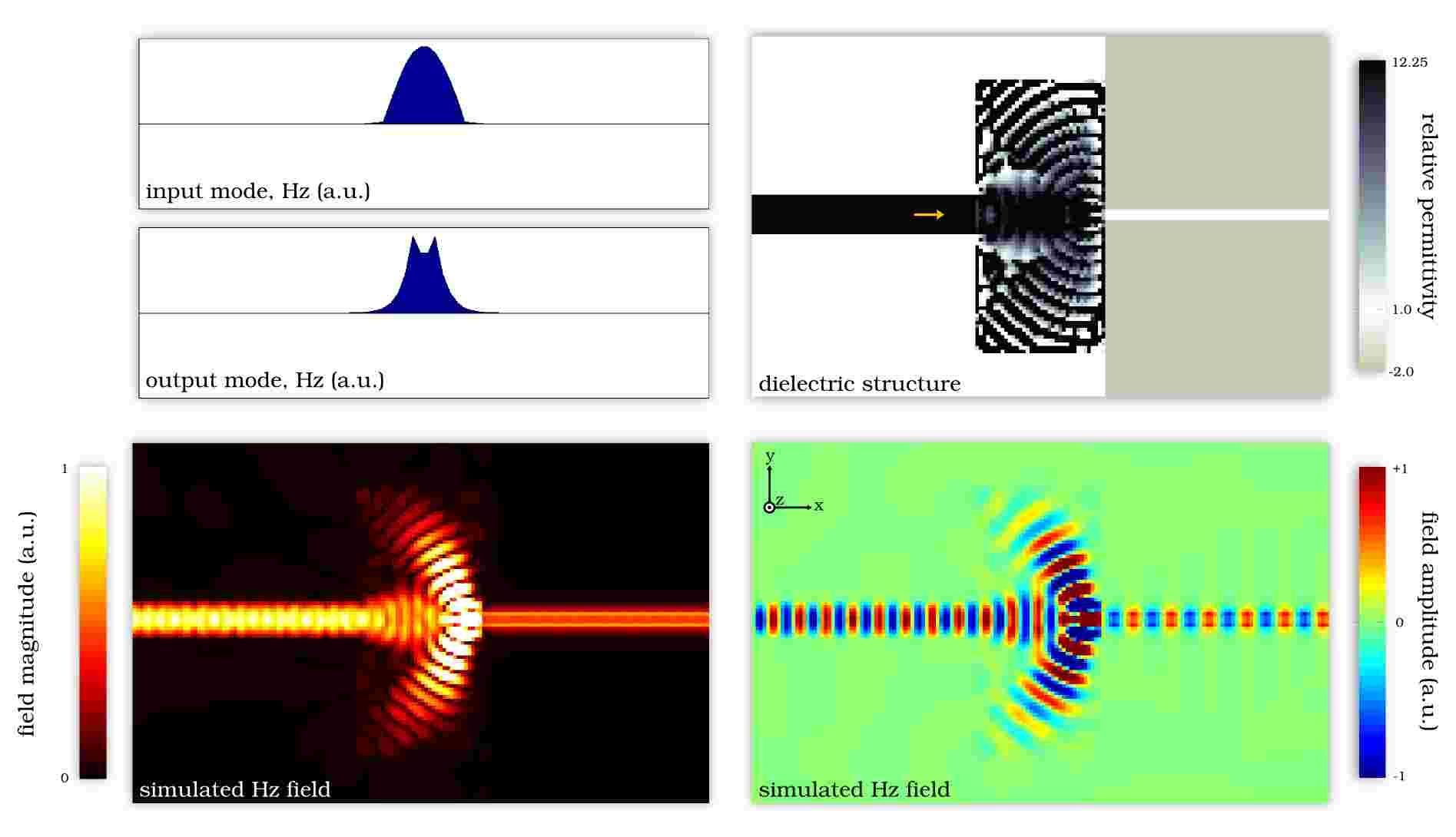}{Coupler to a plasmonic metla-insulator-metal waveguide.
            Efficiency: 97.5\%, 
            device footprint: $36 \times 76$ grid points, 
            wavelength: 25 grid points.}

In this case, the permittivity of the metal ($\epsilon = -2$) is chosen to be 
    near the plasmonic resonance ($\epsilon = -1$).

\subsection{Coupling to a metal wire plasmonic waveguide mode}
\myfig{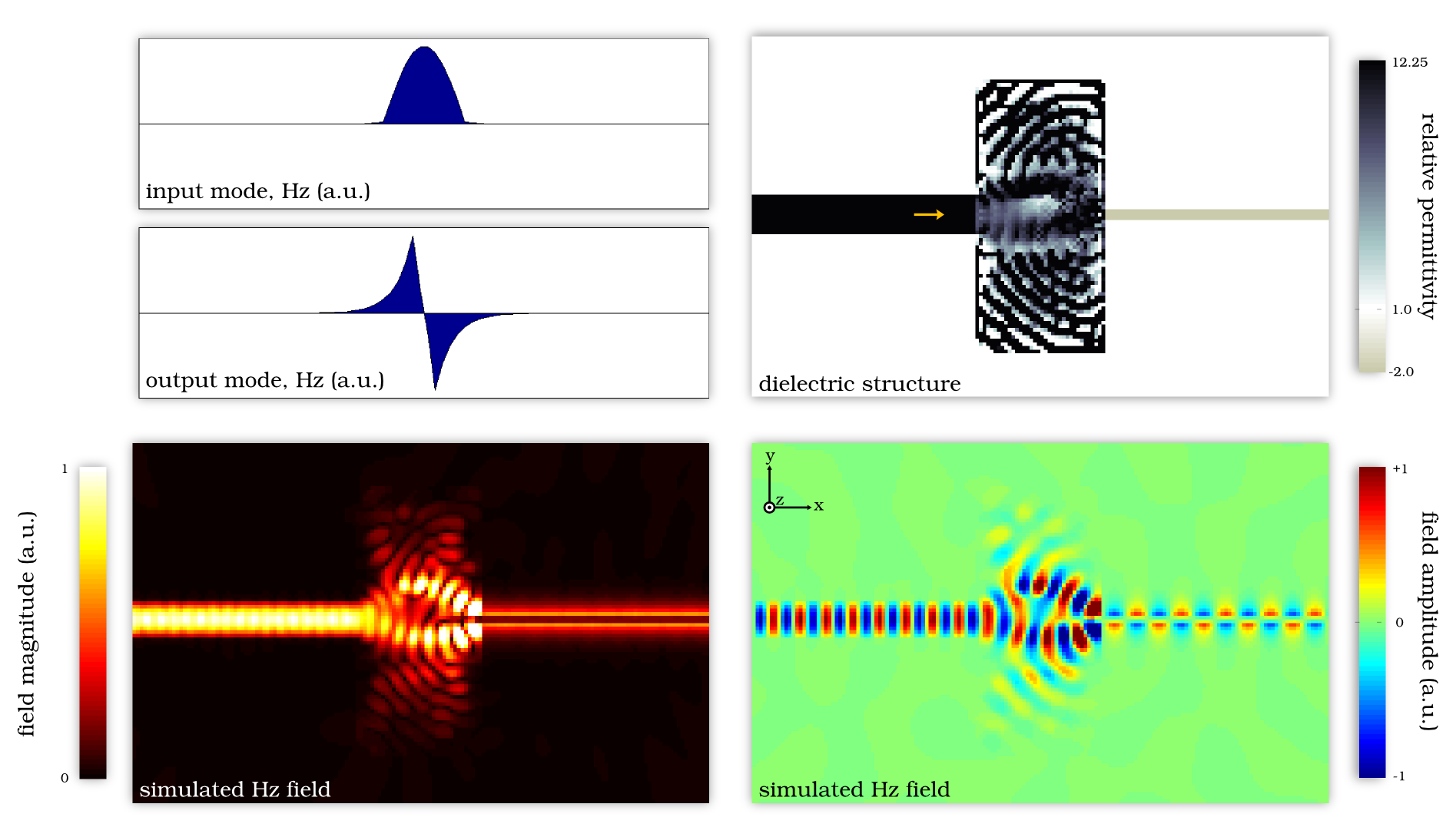}{Coupler to a plasmonic wire waveguide.
            Efficiency: 99.1\%, 
            device footprint: $36 \times 76$ grid points, 
            wavelength: 25 grid points.}
Lastly, \fR{wg/5} shows that efficiently coupling to a plasmonic wire
    is achievable as well.

\section{Optical cloak design}
In the previous section,
    we showed that couplers between virtually any two waveguide modes
    could be constructed using the objective-first design method,
    and based on the generality of the method
    one can guess that it may also be
    able to generate designs for any linear nanophotonic device.

Now, we extend the applicability of our method
    to the design of metamaterial devices which operate in free-space.
In particular,
    we adapt the waveguide coupler algorithm to the 
    to the design of optical cloaks.

\subsection{Application of the objective-first strategy}
Adapting the method used in \sR{wg} to the design of optical cloaks
    really only requires one to change the simulation environment
    to allow for free-space modes.
This is accomplished by modifying the upper and lower boundaries
    of the simulation domain from absorbing boundary conditions
    to periodic boundary conditions,
    which allows for plane-wave modes to propagate without loss
    until reaching the left or right boundaries,
    where absorbing boundary conditions are still maintained.

In terms of the design objective, 
    we allow the device to span the entire height of the simulation domain,
    and thus consider only the leftmost and rightmost planes as boundary values.
Specifically, for this section the input and output modes are plane waves
    with normal incidence, 
    as can be expected for good cloaking devices.
The achieved results all yield high efficiency,
    although we note that the cloaking effect is only measured
    for a specific input mode.
That is to say, just as the waveguide couplers previously designed
    were single-mode devices,
    so the cloaks designed in this section are also ``single-mode'' cloaks.

An additional modification, as compared to \sR{wg}, is that
    we now disallow the structure to be modified in certain areas
    which, naturally, contain the object to be cloaked.

With these simple changes we continue to solve \ER{ob1}
    with the alternating directions method
    in order to now design optical cloaks
    instead of waveguide couplers.
Once again, as in \sR{wg}, each design is run for 400 iterations
    with a uniform initial value of $p = 1/9$ for the structure
    (where the structure is allowed to vary),
    and the range of $p$ is limited to $1/12.25 \le p \le 1$,
    implying a dielectric cloak.

\subsection{Anti-reflection coating}
As a first example,
    we attempt to design the simplest and most elementary ``cloaking'' device available,
    which, we argue, is a simple anti-reflection coating;
    in which case the object to be cloaked is nothing more than
    the interface between two dielectric materials.
In this case we use the interface between air and silicon, as shown in \fR{cloak/c1}
\myfig{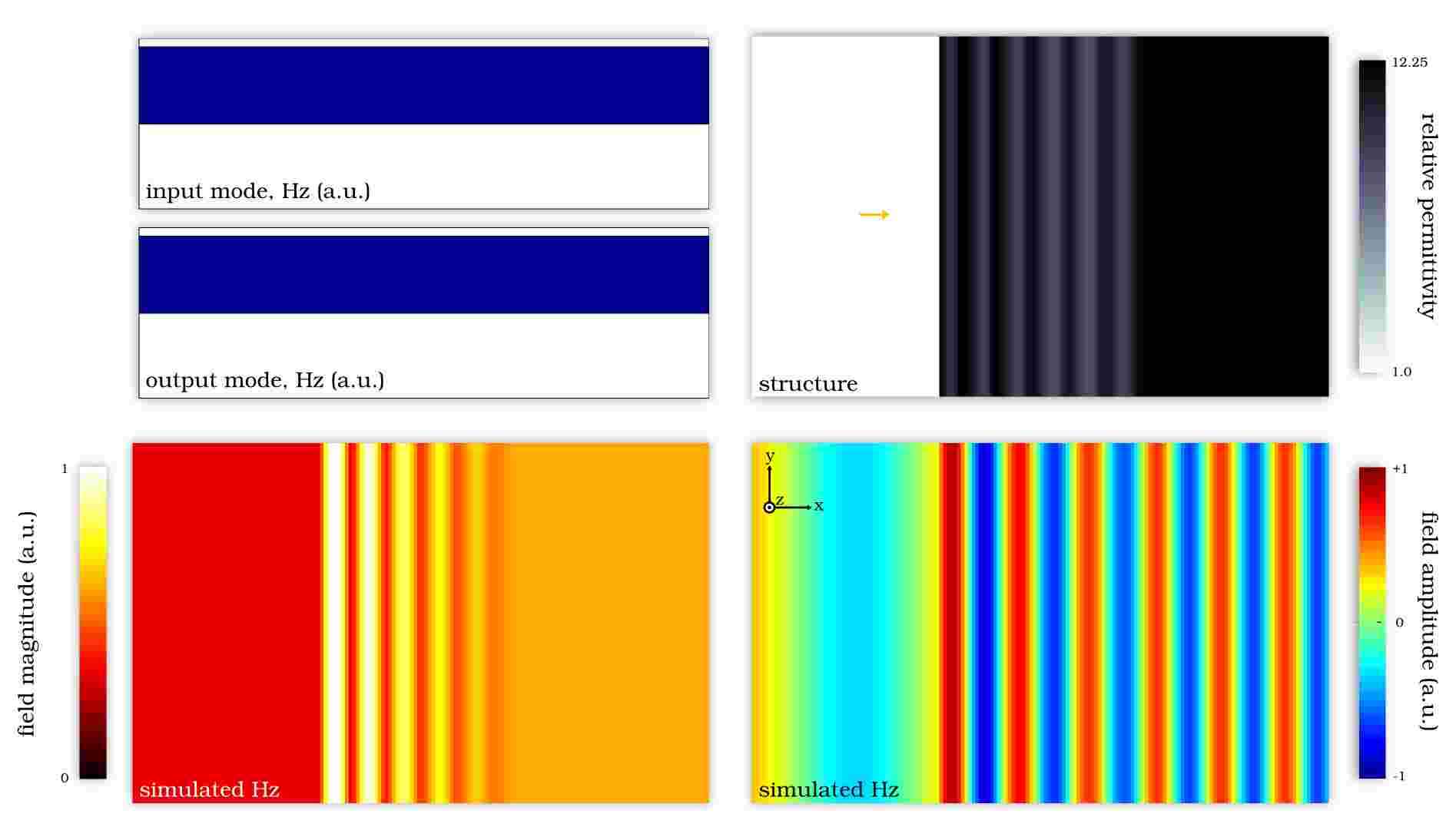}{Anti-reflection coating.
                Efficiency: 99.99\%,
                device footprint: $60 \times 100$ grid points,
                wavelength: 63 grid points.}

Unsurprisingly for such a simple case, 
    we achieve a very high efficiency device.
Note also that the efficiency of the device can be deduced by eye,
    based on the absence of reflections or standing waves 
    in bottom two plots of \fR{cloak/c1}.

\subsection{Wrap-around cloak}
Next, we design a cloak for a plasmonic cylinder,
    which is quite effective at scattering light
    as can be seen from \fR{cloak/c6}.
\myfig{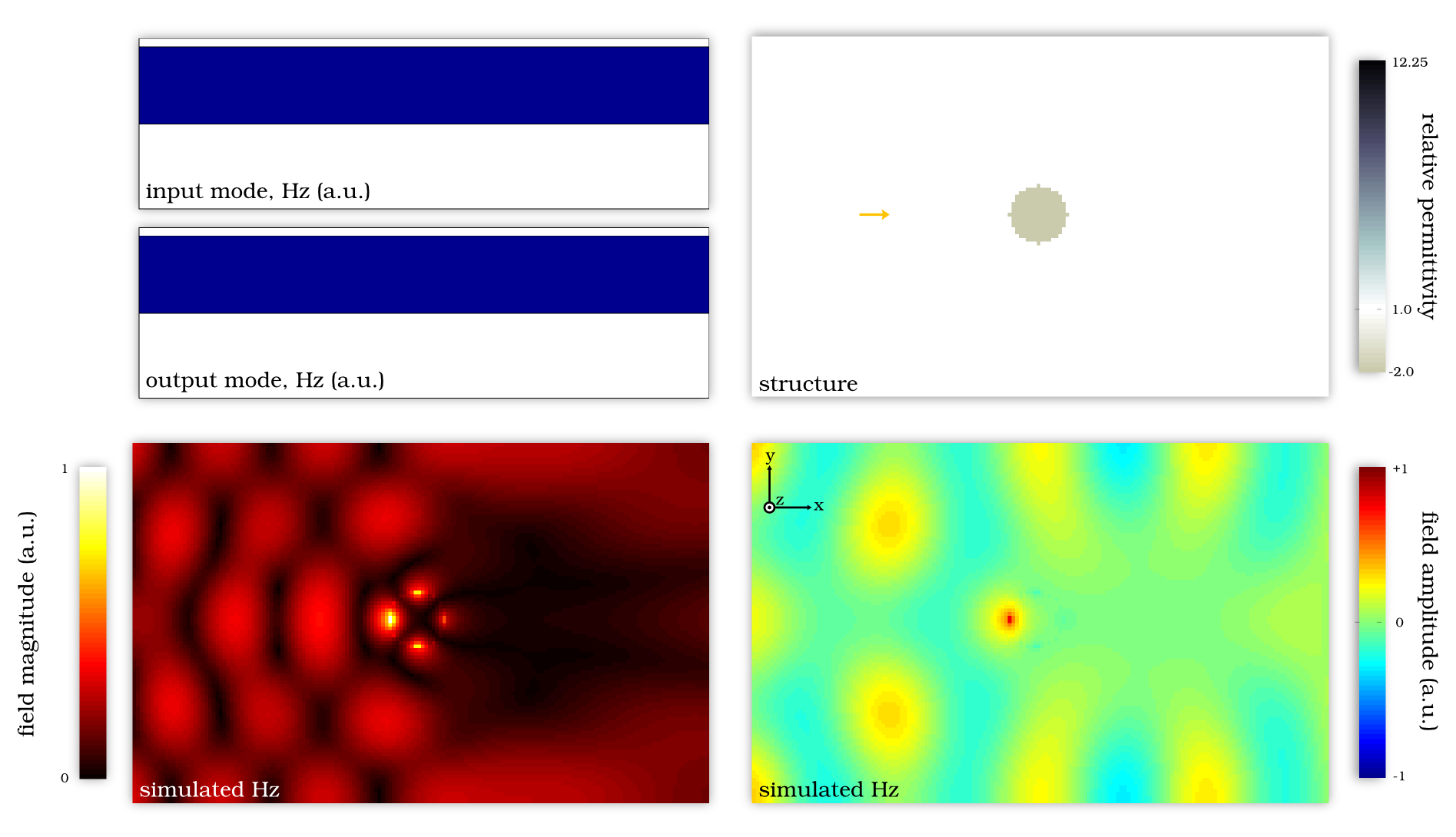}{Plasmonic cylinder to be cloaked. 
                68.5\% of light is diverted away from the desired output mode.}

In designing the wrap-around cloak,
    we allow the structure to vary at all points within the design area
    except in the immediate vicinity of the plasmonic cylinder.
Application of the objective-first strategy results
    in an efficient device as seen in \fR{cloak/c2}.
\myfig{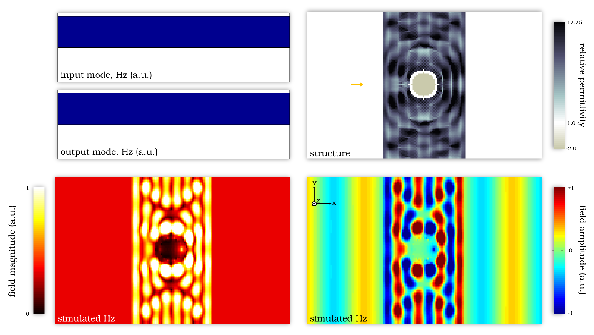}{Wrap-around cloak.
                Efficiency: 99.99\%,
                device footprint: $60 \times 100$ grid points,
                wavelength: 42 grid points.}

Note that our cloak employs only isotropic, non-magnetic materials,
    and at the same time it is specific to a particular input
    and to a particular object.

\subsection{Open-channel cloak}
With a simple modification, from the previous section,
    we can design a cloak which features an open channel
    to the exterior electromagnetic environment.
This simple modification is forcing an air tunnel
    to be opened which connects the cylinder to the outside world
    both toward its front and back.
\myfig{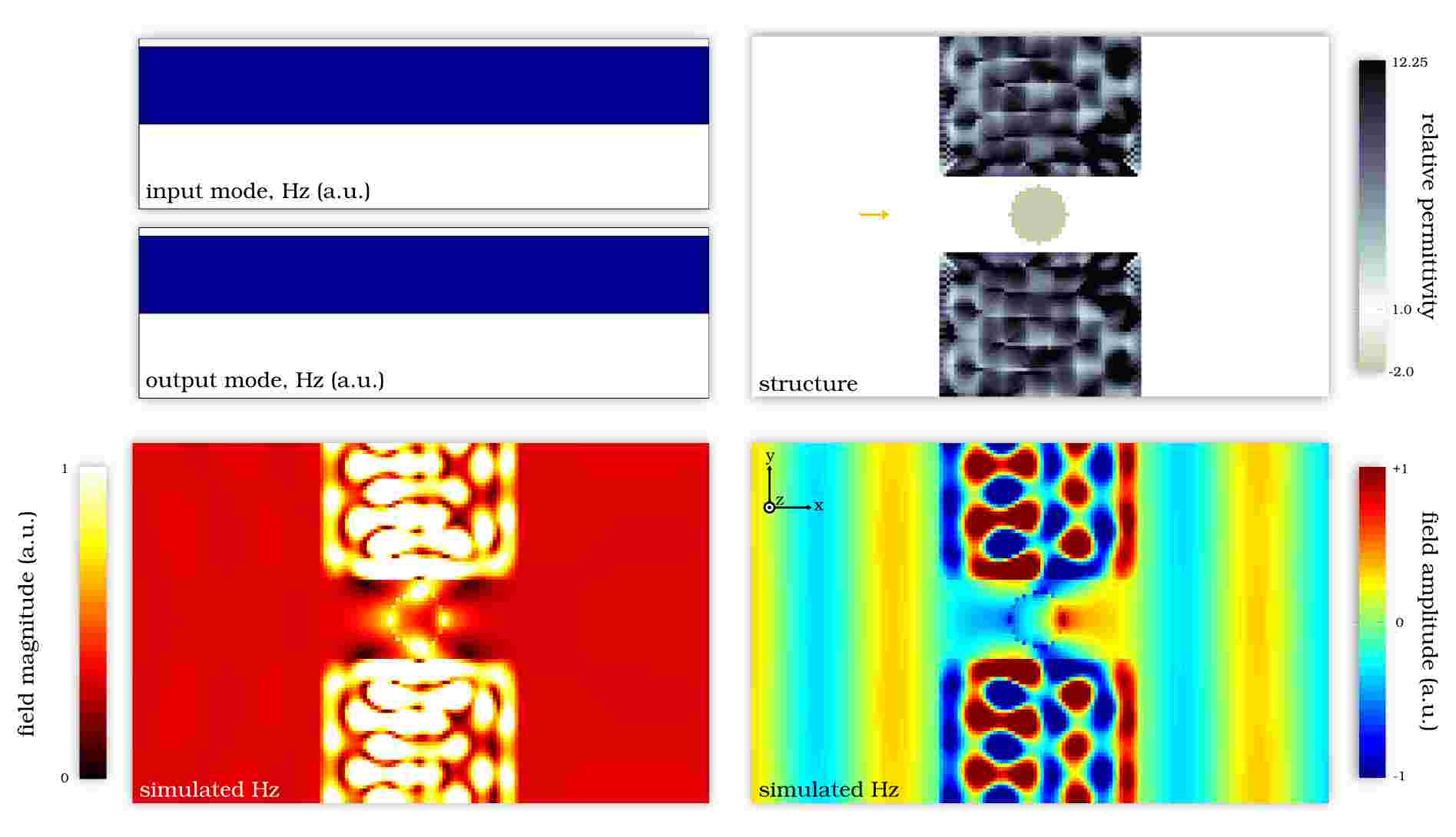}{Open-channel cloak.
                Efficiency: 99.8\%,
                device footprint: $60 \times 100$ grid points,
                wavelength: 42 grid points.}

Such a design is still very efficient
    and exhibits the usefulness of the objective-first strategy
    in cases where other methods, such as transformation optics,
    may not be able to be applied.

\subsection{Channeling cloak}
Our last cloaking example replaces the plasmonic cylinder 
    with a thin metallic wall in which a sub-wavelength channel is etched.
Such a metallic wall is very effective at blocking incoming light
    (as can be seen from \fR{cloak/c7}) 
    because of its large negative permittivity ($\epsilon = -20$),
    meaning that any cloaking device would be forced to channel
    all the input light into a very small aperture
    and then to flatten that light out into a plane wave again.
\myfig{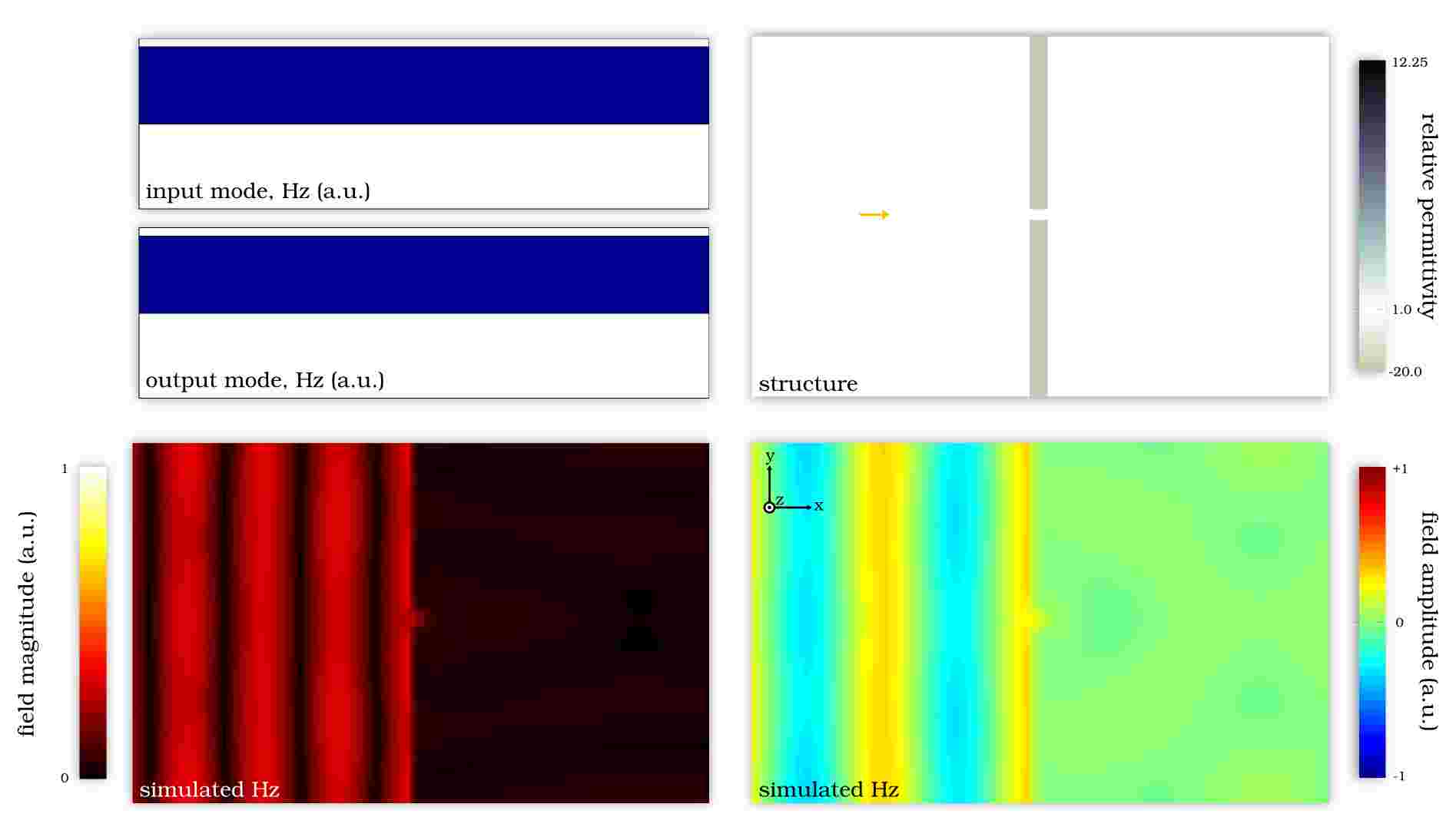}{Metallic wall with sub-wavelength channel to be cloaked.
                99.9\% of the light is blocked from the desired output plane-wave.}

Once again, our method is still able to produce a very efficient design,
    as shown in \fR{cloak/c5}.
\myfig{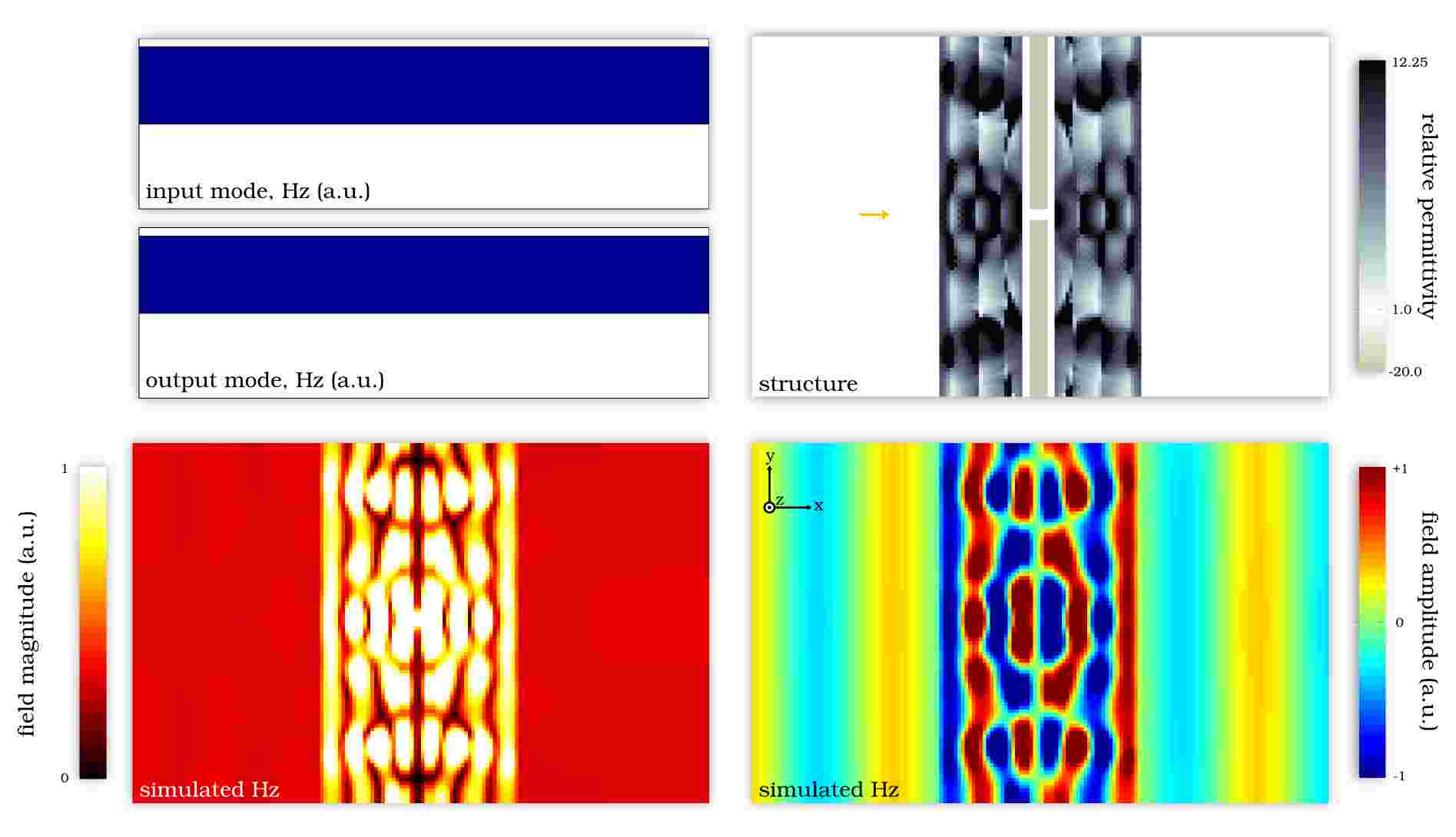}{Channeling cloak.
                Efficiency: 99.9\%,
                device footprint: $60 \times 100$ grid points,
                wavelength: 42 grid points.}

\section{Optical mimic design}
We now apply our objective-first strategy to the design of optical mimics.

We define an optical mimic to be a linear nanophotonic device which 
    mimics the output field of another device.
In this sense optical mimics are the anti-cloaks;
    whereas cloaks strive to make an object's electromagnetic presence vanish,
    mimics strive to implement an object's presence without that 
    object actually being there.

As such, the design of optical mimics provides a tantalyzing approach
    to the realization of practical metamaterial devices.
That is to say, if one can reliably produce practical optical mimics,
    then producing metamaterials (which are often based on fictitious materials)
    can be accomplished by simply producing an optical mimic of that material.

In a more general sense, 
    designing optical mimics is really just a recasting of the thrust of 
    the objective-first design strategy in its purest form:
    the design of a nanophotonic device based purely 
    on the electromagnetic fields one wishes to have it produce.
As such, devices which perform well-known optical functions
    (e.g. focusing, lithography) can also be designed.

\subsection{Application of the objective-first strategy}
The objective-first design of optical mimics proceeds in virtually
    an identical way to the design of optical cloaks,
    the only difference being that the output modes are 
    specifically chosen to be those which produce the desired function.
For most of the examples provided, the input illumination is still an incident plane wave.

Lastly, instead of measuring efficiency, 
    we measure the relative error of the simulated field
    against that of a perfect target field
    at a relevant plane of some distance away from the device.
The location of this plane is identified as a dotted line in the subsequent figures.

\subsection{Plasmonic cylinder mimic}
Our first mimic is simply to mimic the plasmonic cylinder 
    which we cloaked in the previous section.
\myfig{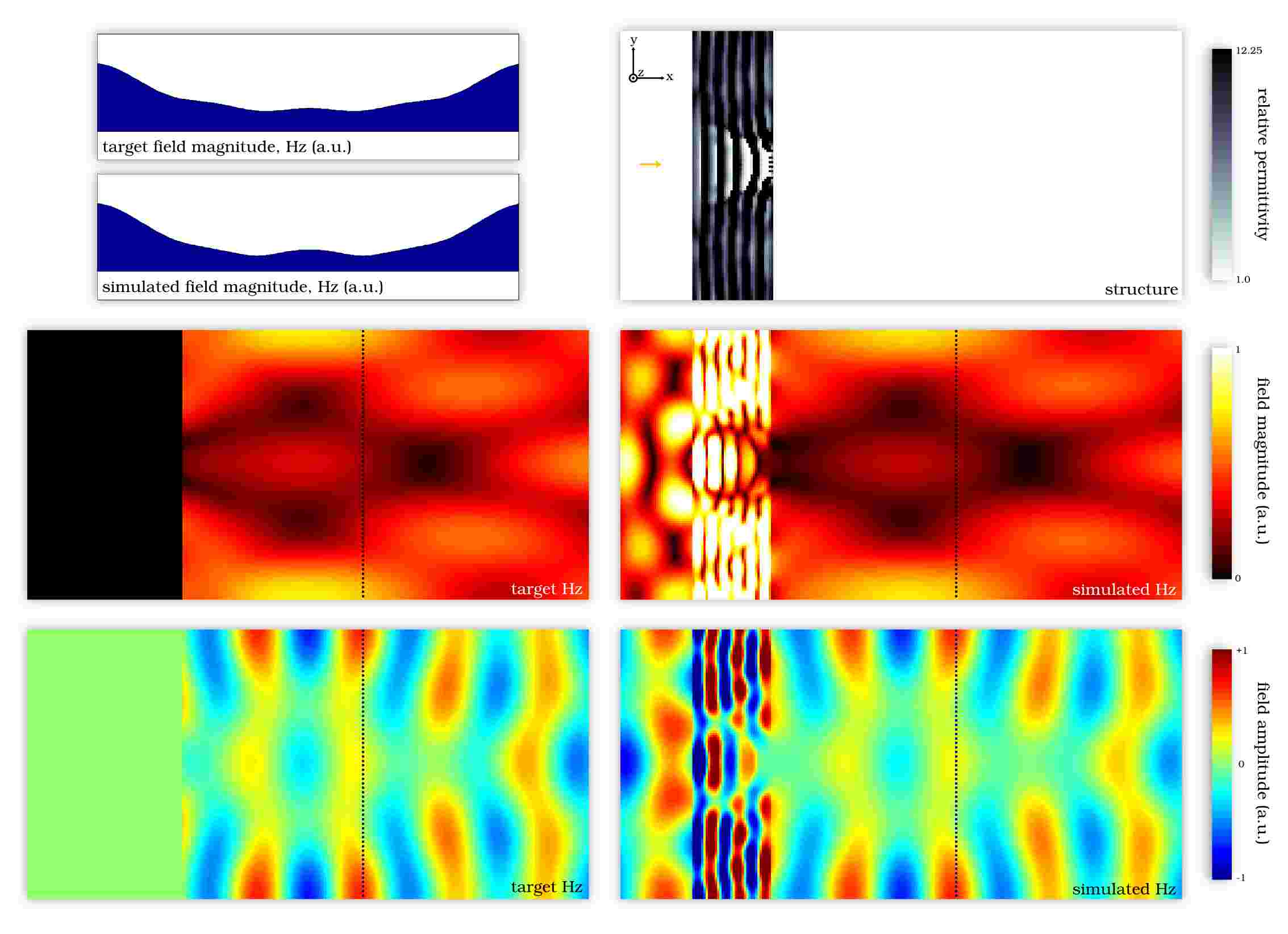}{Plasmonic cylinder mimic (see \fR{cloak/c6} for the original object). 
                Error: 8.1\%,
                device footprint: $40 \times 120$ grid points,
                wavelength: 42 grid points.}

\FR{mimic/m1} shows the result of the design.
The final structure is shown in the upper right plot,
    while the ideal field and the simulated field
    are shown in the middle and bottom plots.
Note that the ideal field is cut off to emphasize 
    the fields to the right of the device (the output fields).
Also, the magnitude of the fields are compared at the 
    dotted black line at which point the relative error
    is also calculated.
For this simple, initial mimic, the simulated field is
    quite closely imitates that produced by a single plasmonic cylinder.

\subsection{Diffraction-limited lens mimic}
We now design a mimic for a typical diffraction-limited lens.
In this case, the object which we wish to mimic does not require simulation
    since the fields of a lens can be readily computed.
For the three figures below, 
    the computed ideal fields are shown as the target fields.

\FR{mimic/m3} shows the mimic of a lens with a moderate focus spot. 
In such a lens, the focusing action is gradual and easily discernable
    by eye.

\myfig{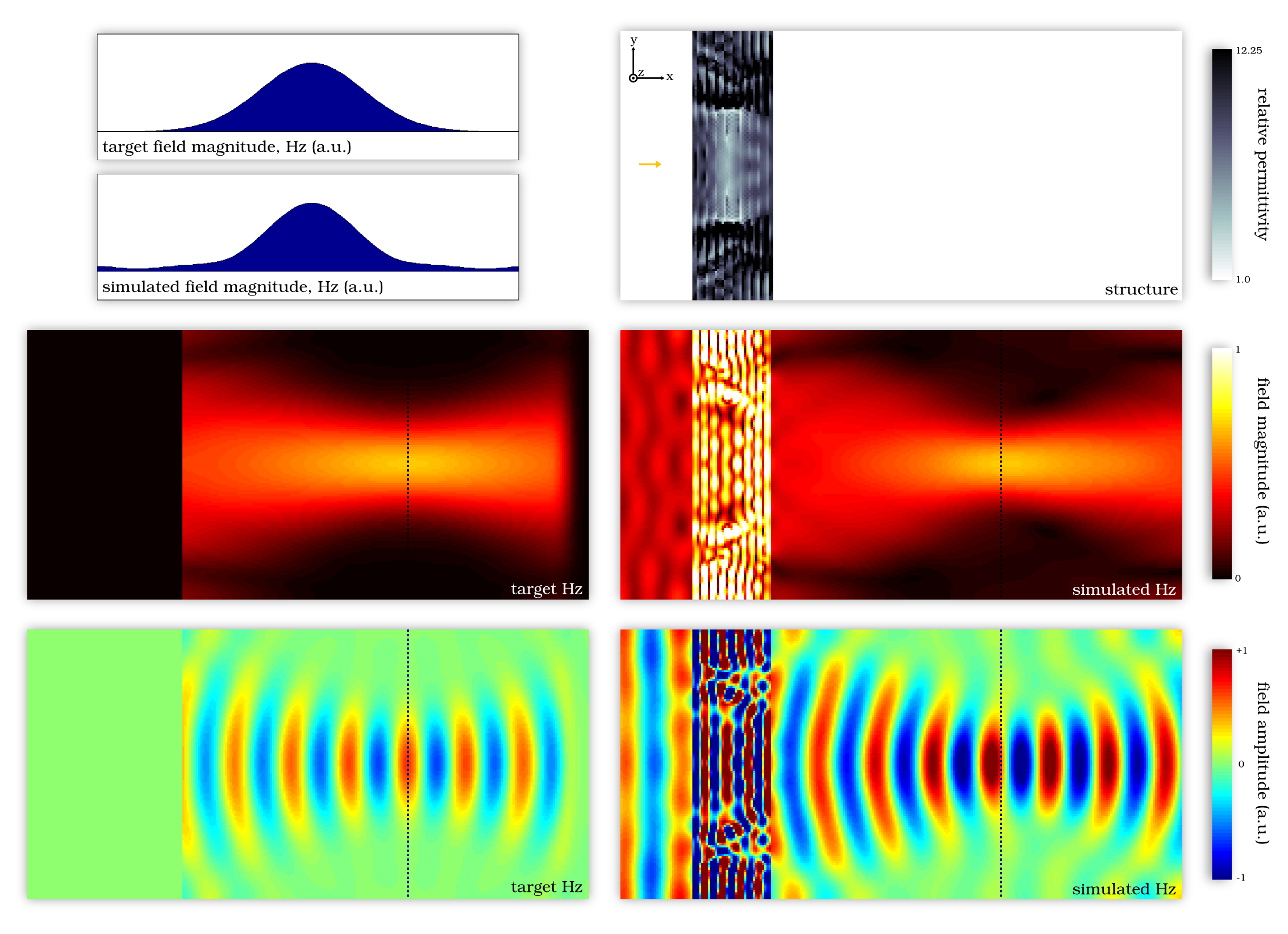}{Full-width-half-max at focus: 1.5 $\lambda$,
                focus depth: 100 grid points.
                Error: 12.0\%,
                device footprint: $40 \times 120$ grid points (1.6 $\lambda$ thick),
                wavelength: 25 grid points.}

In constrast, \fR{mimic/m4} and \fR{mimic/m5} are both mimics of
    a lens with a smaller half-wavelength spot size.
Such a lens is much harder to design, 
    because of the high-frequency spatial components involved;
    and yet, we show that an objective-first strategy can 
    produce successful designs with both smaller and larger focus depths.
\myfig{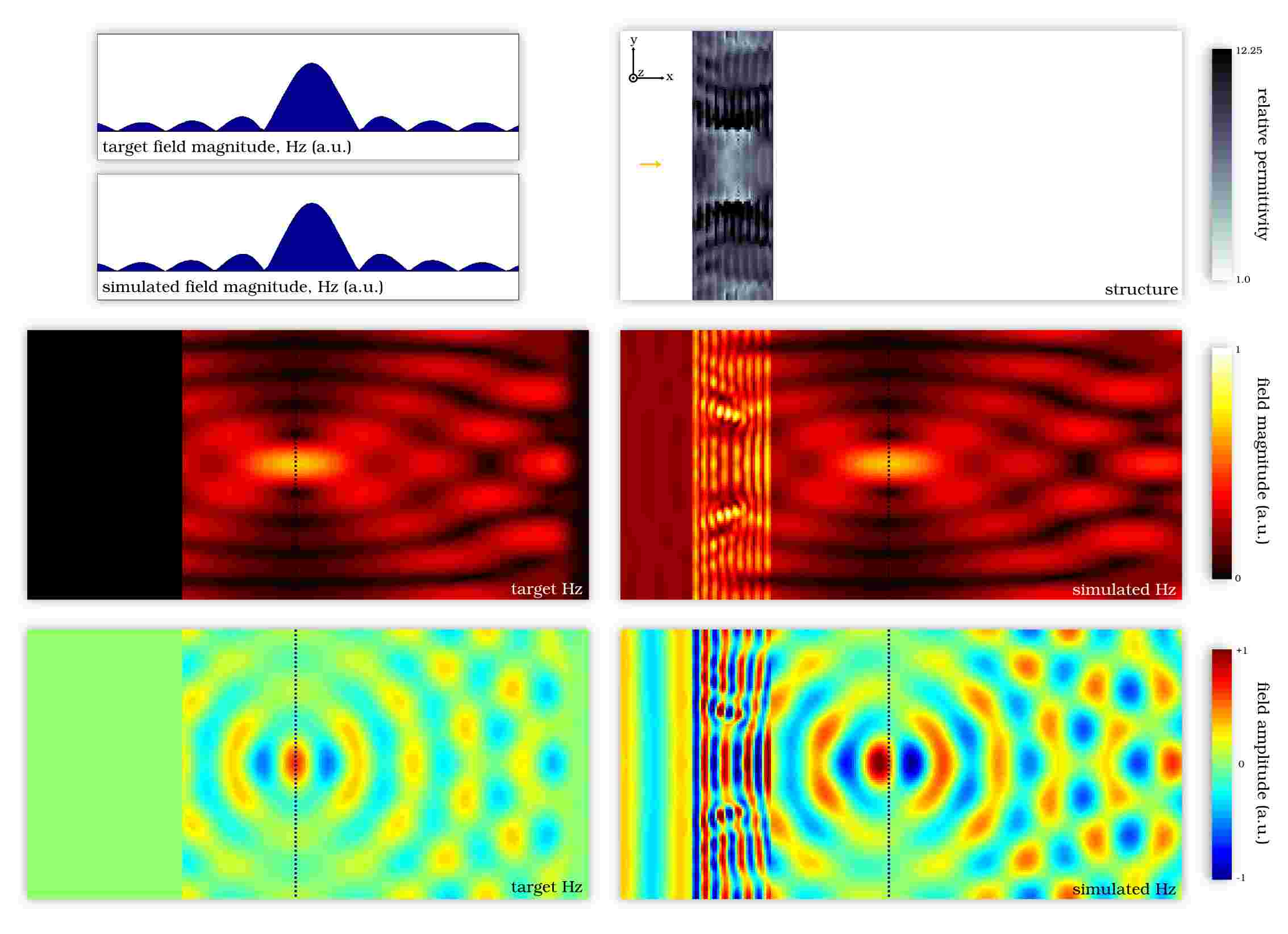}{Full-width-half-max at focus: 0.5 $\lambda$,
                focus depth: 50 grid points.
                Error: 5.6\%,
                device footprint: $40 \times 120$ grid points (1.6 $\lambda$ thick),
                wavelength: 25 grid points.}
\myfig{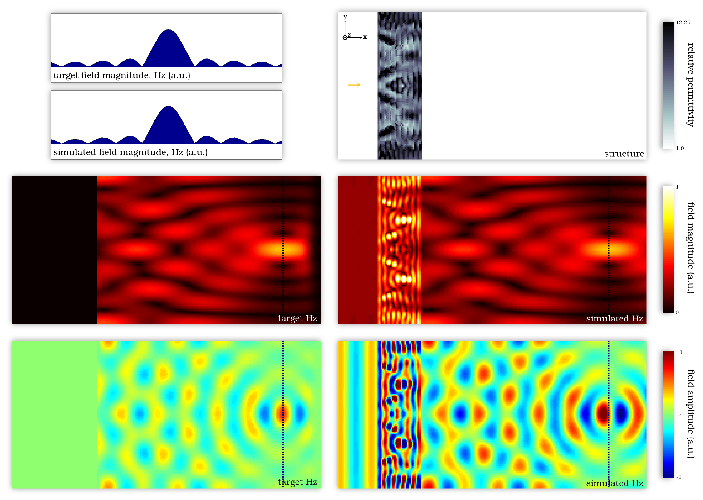}{Full-width-half-max at focus: 0.5 $\lambda$,
                focus depth: 150 grid points.
                Error: 1.4\%,
                device footprint: $40 \times 120$ grid points (1.6 $\lambda$ thick),
                wavelength: 25 grid points.}
\clearpage

\subsection{Sub-diffraction lens mimic}
Our method is now employed to mimic the effect of a sub-diffraction lens.
Since such a lens can be created using a negative-index material
    this mimic can be viewed as an imitation of a negative-index material,
    in that the following device recreates the sub-diffraction target-field 
    at the output plane (dotted line)
    when illuminated by the same target field at the input of the device.
In other words,
    this device is an image-specific sub-diffraction imager,
    which is another way of saying that it is a single-mode imager.
\myfig{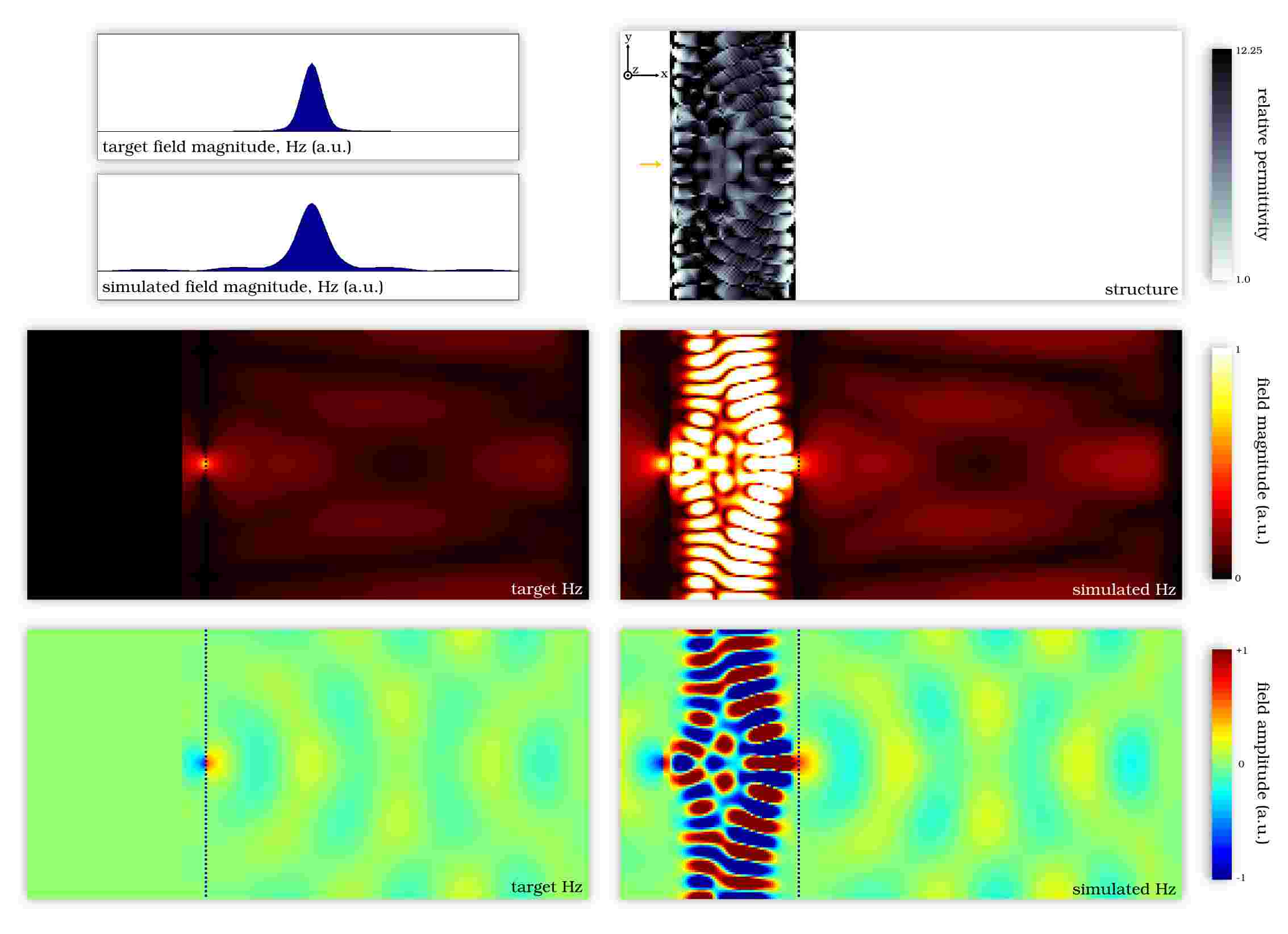}{Sub-diffraction lens mimic.
                The target field has a full-width half-maximum of 0.14 $\lambda$.
                Error: 28.6\%,
                device footprint: $60 \times 120$ grid points (1.43 $\lambda$ thick),
                wavelength: 42 grid points.}

As \fR{mimic/m2} shows,
   we are able to recreate the target field at the output.
Note that the target field is created simply by placing 
    the imaging field at the output plane.
Also note that, as expected, 
    the output field decays very quickly since,
    for such a deeply subwavelength field,
    it is composed primarily of evanescently decaying modes.

\subsection{Sub-diffraction optical mask}
Lastly, we extend the idea of a sub-diffraction lens mimic
    one step further and
    design a sub-diffraction optical mask.
Such a device takes a plane wave as its input and
    produces a sub-diffraction image at its output plane.
Of course, akin to its lens counterpart,
    this output plane must lie within the near-field 
    of the device (specifically, two computational cells away)
    because of its sub-wavelength nature.
\myfig{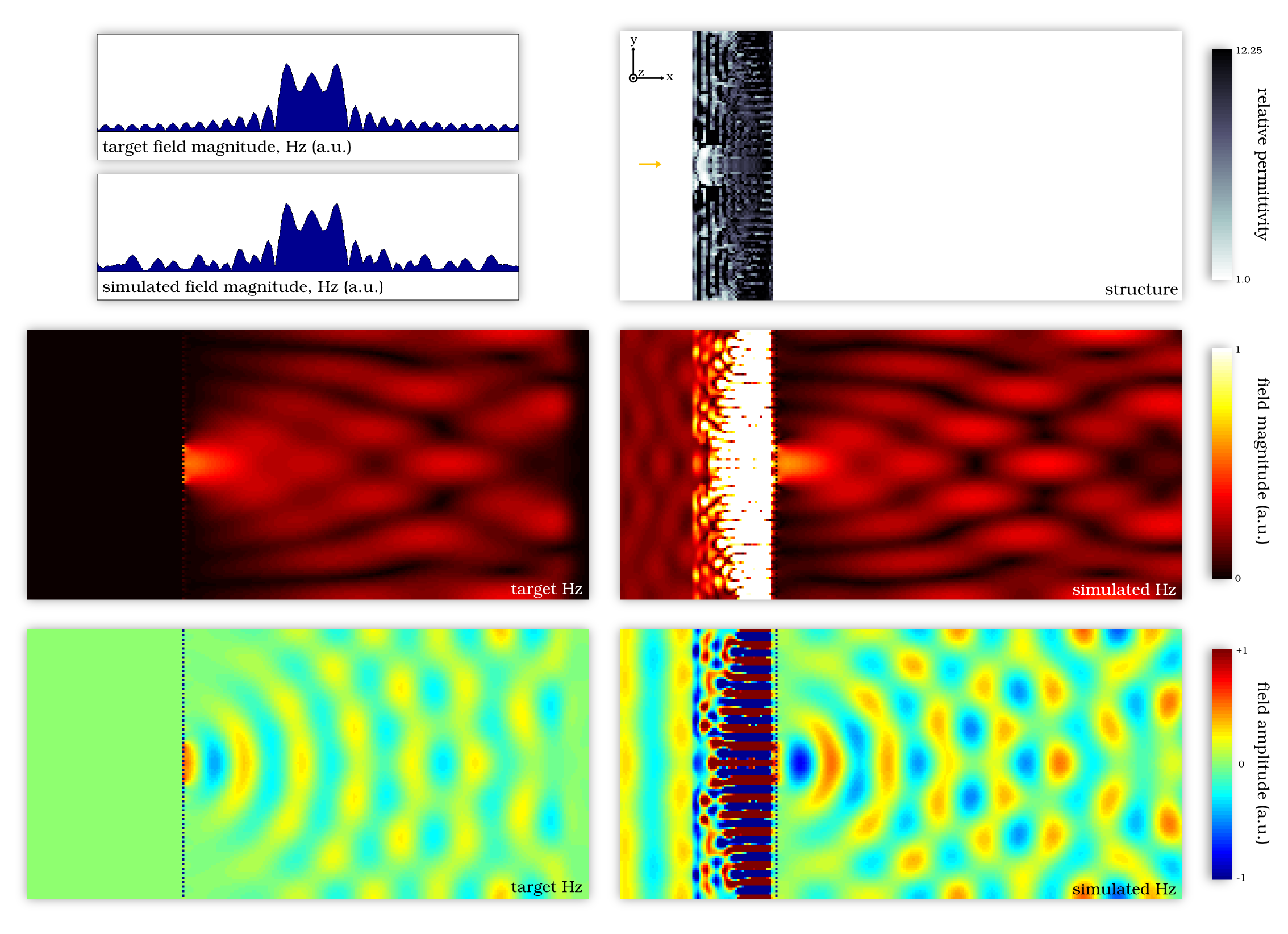}{Sub-diffraction optical mask.
                The three central peaks in the target field are each
                separated by 0.28 $\lambda$.
                Error: 19.8\%,
                device footprint: $40 \times 120$ grid points,
                wavelength: 25 grid points.}

\FR{mimic/m6} shows the design of a simple mask which 
    successfully produces three peaks at its output.

\section{Extending the method}
The objective-first method, as applied in the examples in this chapter,
    represents only a small foray into the area of nanophotonic design.
Several key extensions to what is presented here are needed to
    fully address real-world nanophotonic design challenges.

\subsection{Three-dimensional design}
The first of these is the need to design fully three-dimensional structures.
Doing so provides no inherent difficulties aside from the matrices in
    \ER{Ab} becoming very large.
This is not insurmountable as electromagnetic simulation software for 
    three-dimensional nanophotonic structures already exists.

In fact, for certain choices of the design objective 
    (i.e. those of low-rank) 
    \ER{Fsub} can be efficiently solved by a small number of calls
    to unmodified simulation software. 
Of course, for general design objectives, such software will need 
    to be modified in order to solve \ER{Fsub}.

On the other hand, specialized software to solve \ER{Bd} in 
    any number of dimensions does not exist,
    although this was not a problem in two dimensions since generic 
    linear algebra solvers are more than accurate.
In three dimensions, 
    the large size of matrix $B(x)$ can be greatly compressed by considering
    only fabrication processes which modify a structure in-plane.
In this way, the degrees of freedom in $p$ can be greatly reduced and the original
    methods used in this chapter can still be applied.
This work-around is especially viable since in-plane structures are 
    of most interest from a practical standpoint.
    
\subsection{Multi-mode}
A second necessary extension is to be able to consider the multiple fields
    that a structure produces in response to input fields of differing frequency
    and spatial distribution.
Such an extension is straightforward in the objective-first formulation
    and results in the following modified problem statement,
\BA \minimize{x_i,p} \sum_i \| A(p) x_i - b(p) \|^2 \notag \\
    \subto f(x_i) = f_{i,\text{ideal}} \quad \text{i = 1, \ldots, n} \\
        & p_0 \le p \le p_1, \notag \EA
    which can be separated into field and structure sub-problems 
    as in the single-mode formulation.
In the multi-mode case, this results in one structure sub-problem 
    and $n$ field sub-problems.
Interestingly, the $n$ field sub-problems lend themselves 
    naturally to parallelization
    since they can be solved independently,
    leading to the possibility that a multi-mode design completing in
    roughly the same time as a single-mode design.

\subsection{Binary structure}
Another necessary extension of our method is to force 
    the values of $p$ to be discrete.
This is not trivial since a naive restatement of \ER{ob1}
    which includes such a constraint,
\BA \minimize{x,p} \| A(p) x - b(p) \|^2 \notag \\
    \subto f(x) = f_\text{ideal} \\
        & p \in \{p_0, p_1\}, \notag \EA
    results in a very difficult combinatorial problem.

Tractable approaches include penalizing intermediate values of $p$ \cite{simp}
    or even transferring to a level-set method \cite{miller}
    where the distinction between materials is explicit.

\subsection{Robustness}
Lastly, the design of structures which are robust to both
    fabrication imperfections and fluctuations in environmental parameters
    is also a necessity for practical real-world devices.

It seems likely in this case that a heurestic approach may
    be most successful in this case, rather than to tackle 
    the problem head-on.
For instance, to account for fluctuating material parameters
    induced by temperature changes
    one may design a device to operate for a larger-than-necessary 
    frequency range.

\section{Conclusion}
We have introduced an objective-first approach
    to the design of nanophotonic components, and
    applied it to the design of waveguide couplers,
    optical cloaks, and optical mimics.
In doing so, we hope to have exhibited both the simplicity
    and the breadth of our method to the design of a broad
    class of linear, single-mode devices.
In addition to posting the source code for all the examples online \cite{code}, 
    we have outlined the necessary extensions to our method in order
    to design practical, three-dimensional devices.

\section{Acknowledgements}
This work has been supported by the 
    AFOSR MURI for Complex and Robust On-chip Nanophotonics 
    (Dr. Gernot Pomrenke), grant number FA9550-09-1-0704; 
    and the Alexander von Humboldt Foundation.

\backmatter


\end{document}